\documentclass[aps]{revtex4}
\usepackage{amsmath}
\usepackage{amsthm}

\newtheorem{theorem}{Theorem}[section]
\newtheorem{corollary}[theorem]{Corollary}

\newtheorem{lemma}[theorem]{Lemma}

\def\Rul#1#2{R{}^{#1}{}_{#2}}
\def\Rulul#1#2#3#4{R{}^{#1}{}_{#2}{}^{#3}{}_{#4}}
\def\Ruull#1#2#3#4{R{}^{#1#2}{}_{#3#4}}

\def\Cuull#1#2#3#4{C{}^{#1#2}{}_{#3#4}}

\def\Ful#1#2{F{}^{#1}{}_{#2}}

\newenvironment{example}{\refstepcounter{theorem}\trivlist
        \item[\hskip \labelsep{\textbf{Example \thetheorem.}}]}
       {\qed\endtrivlist}

\newif\ifleftleg
\newif\ifrightleg
\newif\ifleftdot
\newif\ifrightdot
\newif\ifsign
\def\legs#1{\begin{picture}(30,15)
                \ifsign
                        \put(15,13){\circle{6}}
                        \put(10,10){\makebox(10,6)[c]{#1}}
                \else
                        \put(5,9){\makebox(20,7)[c]{#1}}
                \fi
                \ifleftleg
                        \ifleftdot
                                \put(9,7){\circle*{0.5}}
                                \put(7,5){\circle*{0.5}}
                                \put(5,3){\circle*{0.5}}
                        \else
                                \ifsign
                                        \put(13,11){\line(-1,-1){11}}
                                        \put(5,7){\makebox(0,0)[c]{$+$}}
                                \else
                                        \put(12,10){\line(-1,-1){10}}
                                \fi
                        \fi
                \fi
                \ifrightleg
                        \ifrightdot
                                \put(21,7){\circle*{0.5}}
                                \put(23,5){\circle*{0.5}}
                                \put(25,3){\circle*{0.5}}
                        \else
                                \ifsign
                                        \put(17,11){\line(1,-1){11}}
                                        \put(25,7){\makebox(0,0)[c]{$-$}}
                                \else
                                        \put(18,10){\line(1,-1){10}}
                                \fi
                        \fi
                \fi
             \end{picture}
}

\begin{document}

\title{Dimensionally Dependent Tensor Identities by Double Antisymmetrisation.}
\date{\today}
\author{S. Brian Edgar}
\email{bredg@mai.liu.se}
\author{A. H\"oglund}
\email{anhog@mai.liu.se}
\affiliation{Matematiska institutionen,
        Link\"opings universitet,
        SE-581 83 Link\"oping,
        Sweden.
}

\begin{abstract}
Some years ago, Lovelock showed that a number of apparently unrelated
familiar tensor identities had a common structure, and could all be
considered consequences in $n$-dimensional space of a pair of
fundamental identities involving trace-free $(p,p)$-forms where
$2p\geq n$.  We generalise Lovelock's results, and by using the fact
that associated with \emph{any} tensor in $n$-dimensional space there
is associated a fundamental tensor identity obtained by
antisymmetrising over $n+1$ indices, we establish a very general
'master' identity for \emph{all} trace-free $(k,l)$-forms. We
then show how various other special identities are direct and simple
consequences of this master identity; in particular we give direct
application to Maxwell, Lanczos, Ricci, Bel and Bel-Robinson tensors,
and also demonstrate how relationships between scalar invariants of
the Riemann tensor can be investigated in a systematic manner.
\end{abstract}

\maketitle

\section{Introduction}\label{ch:idanti}

In an $n$-dimensional space any tensor expression $T_{a_1a_2 \ldots a_k}$
with $k>n$  indices satisfies a tensor identity
\begin{equation}\label{eq:anti}
        T_{[a_1 a_2...a_k]}=0.
\end{equation}
Such mathematically obvious identities can be very useful in practical
calculations.  However, the antisymmetrisation process may not be so
explicit since it need not be applied only on free indices; it could
also involve dummy indices, some of which could be absorbed into
traces, as we shall demonstrate in the following example.
(Here, and in the rest of this paper, we use the abstract index
notation~\cite{penrose1}.  However, any index notation would probably
work well as long as the above property holds.)
\begin{example}\label{ex:xu}
Dianyan Xu~\cite{xu87} constructed the following two scalar identities
for the Riemann tensor in \emph{four dimensional} spaces,
\begin{equation}\label{eq:xu1}
        \Rul ab R^{bcde} R_{acde}
        =
        \frac{1}{4} R R^{abcd} R_{abcd}
        +2 R^{ac} R^{bd} R_{abcd}
        +2 \Rul ab \Rul bc \Rul ca
        -2 R \Rul ab \Rul ba
        +\frac{1}{4} R^3
\end{equation}
and
\begin{equation}\label{eq:xu2}
        \Rulul acbd \Rulul cedf \Rulul eafb
        =
        \frac{1}{2} \Ruull abcd \Ruull cdef \Ruull efab
        -\frac{3}{8} R \Ruull abcd \Ruull cdab
        -3 R^{ac} R^{bd} R_{abcd}
        -4 \Rul ab \Rul bc \Rul ca
        +\frac{9}{2} R \Rul ab \Rul ba
        -\frac{5}{8} R^3.
\end{equation}
These were obtained after  a lengthy calculation  by decomposing  the Riemann
tensor $R_{abcd}$ into its Weyl and Ricci components, and then  {\it
using  spinor methods}; in fact Dianyan Xu claimed that it was not possible to
obtain these identities from  the algebraic properties of the Riemann 
tensor alone.

However, it was pointed out by
Harvey~\cite{harvey95} that the first identity could be  obtained directly by
expanding
\begin{equation}\label{eq:h1}
        \Ruull ab{[a}b \Ruull cdcd \Rul e{e]} = 0
\end{equation}
and the second identity by expanding
\begin{equation}\label{eq:h2}
        \Ruull ab{[a}b \Ruull cdcd \Ruull efe{f]}
        = 0
\end{equation}
and combining the result with the first identity.

So it is now clear that such identities exist not only for Riemann
tensors, nor even just for Riemann candidates (any 4-tensors with the
index symmetries of a Riemann tensor); they are in fact valid for any
tensors which permit the antisymmetrisation constructions (4) and (5).

Harvey's approach also confirmed that both identities were valid in
four and lower dimensions --- irrespective of the signature of the
space (a fact that was missing from Dianyan Xu's spinor
derivation). So this insight into the structure of the two identities
not only led to a better appreciation of the relevance of these
identities in the original context (concerning counterterms in
Lagrangians~\cite{jack85,jack87,harvey95}), but also
highlights the importance of identities built from the
antisymmetrisation property~\eqref{eq:anti} in the study of Riemann
scalar invariants.

Of course the pattern and discussion above suggest how to obtain many
more analogous identities for other tensors, for higher dimensions,
and for higher order. In fact the identity~\eqref{eq:xu2} can be
obtained more directly and compactly by instead using the Weyl tensor
$C_{abcd}$ (the trace-free part of the Riemann tensor) in the identity
$\Cuull ab{[a}b \Cuull cdcd \Cuull efe{f]} =
0$~\cite{jack87,harvey95,F}.
\end{example}

It is also interesting to note that it is not just  scalar identities 
which can be
constructed in this manner; for example, similar tensor identities 
can be considered to
underlie the familiar Cayley-Hamilton theorem:

\begin{example}\label{ex:CH}
Antisymmetrising over $n+1$ indices  for matrices $M^a{}_b$ in $n$ dimensions
gives the Cayley-Hamilton theorem
\begin{equation}
        M^{c_1}{}_{[c_1}M^{c_2}{}_{c_2}\ldots
          M^{c_n}{}_{c_n}\delta_{a]}^b
        = 0.
\end{equation}

Considering the Cayley-Hamilton theorem from this viewpoint suggests
generalisations involving more than one matrix,
\begin{equation}
        M^{c_1}{}_{[c_1}N^{c_2}{}_{c_2}\ldots
          P^{c_n}{}_{c_n}\delta_{a]}^b
        = 0.
\end{equation}
This generalised Cayley-Hamilton theorem has been used to find
relations, (syzygies), between scalar invariants of matrices involving
more than one matrix~\cite{sneddon96,sneddon98,ouchterlony97}.
\end{example}

An important question is whether this technique of antisymmetrising is
just a useful ``trick'' in very special circumstances, or whether
there is deeper structure to be better understood and more fully
exploited.

Some time ago Lovelock~\cite{lovelock70} noted the significance of
certain types of identities --- which he called \textbf{dimensionally
  dependent identities} --- and demonstrated their existence and
importance in quite a wide context (Lovelock~\cite{lovelock70} has
defined a dimensionally dependent identity as an identity which is a
direct consequence of the dimension of the space taking on a
particular value, and which therefore is not valid for arbitrary
dimension $n$ in general).  By showing precisely how and where
dimensionality plays its role in familiar identities, Lovelock
revealed a technique which could be generalised to arbitrary
dimension.  In fact Lovelock's technique simply involved the
antisymmetrising process being applied in $n$ dimensions, to tensors
with \emph{two} sets of indices (upper and lower, in practice) and
specialised in two theorems to trace-free tensors with an equal number
$p$ of upper and lower indices where $2p\geq n$.

Lovelock's investigations were motivated mostly by familiar four
dimensional identities satisfied by Weyl tensors (and Weyl candidates,
i.e.  tensors with the algebraic symmetries of the Weyl tensor);
however, there are other identities which appear to be of a similar
nature, but which cannot be confirmed by Lovelock's two theorems.  It
is the purpose of this paper to investigate dimensionally dependent
identities in a systematic manner and obtain a more complete picture;
in doing so we develop results more general than Lovelock's, and
demonstrate that these new results can be used to confirm identities
which cannot be obtained from Lovelock's results, and also to confirm
other identities which can only be confirmed indirectly from
Lovelock's results.

In the next section we make some general observations and present some
examples as further motivation for the subsequent sections.  We
summarise and illustrate Lovelock's results~\cite{lovelock70} in
Section \ref{ch:ldimdepid}. Lovelock's theorems applied only to
trace-free $(p,p)$-forms in $n$ dimensions where $2p\geq n$.  However,
in Section \ref{ch:basth}, we shall develop more general results in
the form of a ``master'' identity which will be applicable to any
tensor in any dimension, and in particular to trace-free $(k,l)$-forms
in any dimension. In Section 5 we will then show that a number of
important identities are consequences of this master identity. In
particular, we show that the simplification of the gravity-matter
coupling terms in the Weyl wave equation and in the Bel tensor, and
the complete symmetry of the super-energy tensor for the Lanczos
potential of the Weyl tensor are all trivial consequences when the
master identity is specialised to four dimensions; in a similar manner
we confirm the complete symmetry of the Bel-Robinson tensor in four
and five dimensions. In addition, we demonstrate how the results
permit a systematic study of relationships between scalar invariants
of the Riemann tensor.  In section~6 we illustrate how the importance
of dimensionally dependent identities have been overlooked by reducing
the algebraic Rainich condition, and point out the potential
usefullness of these results in more general situations.

\section{Identities by Double Antisymmetrisation.}\label{ch:dimdepid}

We consider the tensor $T^{\cal{A}}{}_{a_1}$ where we
have adopted the convention that $\cal{A}$ denotes an arbitrary number
of additional lower and/or upper indices~\cite{penrose1}.

Associated with this tensor $T^{\cal{A}}{}_{a_1}$, in
$n$-dimensional space, there will always be an identity
\begin{equation}\label{eq:id2}
        T^{\cal{A}}{}_{[a_1}\delta_{a_2}^{b_2} \delta_{a_3}^{b_3} \ldots
        \delta_{a_{n+1}]}^{b_{n+1}}
        = 0.
\end{equation}
For future reference  we should note that such identities cannot be 
made 'simpler', in
the sense that  taking the trace of~\eqref{eq:id2} (on explicit 
indices, i.e., indices
which are not implicit in ${\cal{A}}$) simply gives zero on the left 
hand side also.

An alternative approach would be to assume the
presence of a volume element $\eta_{a_1 a_2 ... a_n}$, and formulate analogous
results  in the
$\eta$ notation making use of duals; but we shall concentrate on developing
results using the $\delta$ notation.

There will of course be one such identity associated with each
separate index on $T$.  We could obtain other identities, with less
deltas, by taking more than one index of $T$ explicitly into the
antisymmetrisation operation, but these new identities would involve
only the antisymmetrised part of $T$ with respect to the explicit
indices.

However, when the tensor $T$ is in fact antisymmetric in a group of
$k$ indices, then we can get an identity involving less deltas than
the original identity
\begin{equation}\label{eq:id3}
        T^{\cal{A}}{}_{[a_1 \ldots a_k}\delta_{a_{k+1}}^{b_{k+1}}
        \delta_{a_{k+2}}^{b_{k+2}} \ldots
        \delta_{a_{n+1}]}^{b_{n+1}}
        = 0,
\end{equation}
but without losing any part of $T$.  Again, it is important to note
that taking the trace of~\eqref{eq:id3}, on explicit indices, simply
gives zero on the left hand side also.

We next consider a tensor with both upper and lower indices, each of
which contain groups of antisymmetric indices, i.e.,
$T^{\cal{A}}{}_{a_1 ...  a_k}{}^{b_1 ... b_l} = T^{\cal{A}}{}_{[a_1
...  a_k]}{}^{[b_1 ... b_l]} $ is a \textbf{$\mathbf{(k,l)}$-form} with 
respect to its
explicit indices; then we have the associated identity
\begin{equation}\label{eq:id4}
T^{\cal{A}}{}_{a_1 ...
a_k}{}^{[b_1 ... b_l}\delta_{a_{k+1}}^{b_{k+1}} \dots
\delta_{a_{n+1}}^{b_{n+1}]} = 0
\end{equation}
when $l\geq k.$
  (The analogous case obtained by antisymmetrising  on the lower 
indices for $k\leq l$
is obvious.)

Such classes of tensors include many familiar tensors (Riemann, Weyl,
Ricci, Maxwell, Lanczos, torsion) and they shall be the main focus of
our investigation in this paper.

It is important to note that, although some traces of~\eqref{eq:id4}
yield the trivial identity, \emph{there are some traces
of~\eqref{eq:id4} which yield a non-trivial identity involving the
trace of $T$}. It is these non-trivial traces which yield Lovelock's
identities, and our generalisations of them.  We illustrate this with
the following example.

\begin{example}\label{ex:Riemann}
We consider the $(2,2)$-form
$R_{ab}{}^{cd}$.

\smallskip
\underbar{In three dimensions:}
$R_{ef}{}^{[gh}\delta^c_a \delta^{d]}_{b} = 0.$

Contracting once over $e$ and $g$ gives,
\begin{equation}
        R_{f[a}{}^{[hc} \delta^{d]}_{b]}
        - R_{ef}{}^{e[h} \delta^c_a \delta^{d]}_{b}
        = 0,
\end{equation}
and then over $h$ and $f$ gives,
\begin{equation}
        R_{ab}{}^{cd}
        -4R_{e[a}{}^{e[c} \delta^{d]}_{b]}
        +R_{ef}{}^{ef} \delta_a^{[c} \delta_b^{d]}
        = 0.
\end{equation}
From this last result follows  the well-known fact that the 
trace-free part of a Riemann
tensor
is identically zero in three dimensions.

\smallskip
\underbar{In four dimensions:}
$R_{gh}{}^{[ij} \delta^d_a \delta^e_b \delta^{f]}_c = 0.$

Contracting once over $g$ and $i$ gives,
\begin{equation}
        3R_{h[a}{}^{[jd} \delta^e_b \delta^{f]}_{c]}
        -2 R_{gh}{}^{g[j} \delta^d_a \delta^e_b \delta^{f]}_c
        = 0,
\end{equation}
and then over $h$ and $j$ gives,
\begin{equation}
        3 R_{[ab}{}^{[de}\delta^{f]}_{c]}
        -6 R_{g[a}{}^{g[d} \delta^e_b \delta^{f]}_{c]}
        + R_{gh}{}^{gh} \delta_a^{[d} \delta_b^e \delta_c^{f]}
        = 0.
\end{equation}
A third contraction gives zero on the left hand side also.

\smallskip
All of the above identities can be expressed in a more concise form 
if  $R_{abcd}$ is
decomposed into trace-free  and trace parts.
\end{example}

We note that certain of these contractions yield identities which, if
we were to encounter them not knowing their source, would seem (to our
surprise) to come from antisymmetrising over $n$ or $n-1$ explicit
indices in $n$ dimensions; on the otherhand, when one takes into
account \emph{all} the terms in each identity and also notes that
these contractions involve antisymmetrising over upper and lower
indices, of course, we would realise that our first judgement was
superficial, and we have a disguised antisymmetrisation over $n+1$
indices. However, we have noted above that the identities exist in
their most concise form when presented in terms of the trace free part
of $R_{abcd}$; so such identities can be even more deceptive
especially when constructed explicitly in terms of a trace-free
tensor.

So we now specialise to the important special situation where the
$(k,l)$-form $T^{\cal{A}}{}_{a_1 ... a_k}{}^{b_1 ... b_l} =
T^{\cal{A}}{}_{[a_1 ...  a_k]}{}^{[b_1 ... b_l]} $ is
\emph{trace-free}, i.e., $T^{\cal{A}}{}_{a_1 a_2... a_k}{}^{a_1 b_2
... b_l}=0 $, and the underlying structure of the resulting identities
become less transparent. To illustrate this we will give three simple
examples involving respectively a trace-free $(2,2)$-form
$W_{ab}{}^{cd}$ (of which the Weyl conformal curvature tensor
$C_{ab}{}^{cd}$ is a special example being a symmetric trace-free
$(2,2)$-form since $C_{abcd}=C_{cdab}$, with the additional property
$C_{[abc]d}=0$), a trace-free $(2,1)$-form $L_{ab}{}^c$ (of which the
torsion and Lanczos potential with the additional property
$L_{[abc]}=0$~\cite{lanczos62} are special examples), and a
trace-free $(1,1)$-form $S^a{}_b$ (of which the trace-free symmetric
Ricci tensor $\tilde R^a{}_{b}$, the trace-free symmetric energy
tensor $\tilde T^a{}_{b}$ and the antisymmetric Maxwell tensor
$F^a{}_{b}$, are special examples).

\begin{example}\label{ex:Weyl}
We now apply the above arguments to the trace-free $(2,2)$-form
$W_{ab}{}^{cd}$.

\smallskip
\underbar{In three dimensions:} 
$W_{ef}{}^{[gh} \delta^c_a \delta^{d]}_b = 0.$

Contracting over $f$ and $h$ gives
\begin{equation}
        W_{e[a}{}^{[gc}\delta^{d]}_{b]} = 0,
\end{equation}
and then over $e$ and $g$ gives
\begin{equation}
        W_{ab}{}^{cd} = 0.
\end{equation}

\smallskip
\underbar{In four dimensions:}
$W_{gh}{}^{[ij} \delta^d_a \delta^e_b \delta^{f]}_c = 0.$

Contracting over $h$ and $j$ gives
\begin{equation}
        W_{g[a}{}^{[id} \delta^e_b \delta^{f]}_{c]} = 0,
\end{equation}
and then over $g$ and $i$ gives
\begin{equation}
        W_{[ab}{}^{[de}\delta^{f]}_{c]} = 0.
\end{equation}
A third contraction gives zero on the left hand side also.

\smallskip
\underbar{In five dimensions:}
$W_{ij}{}^{[kl} \delta^e_a \delta^f_b \delta^g_c \delta^{ h]}_d = 0.$

Contracting over $j$ and $l$ gives
\begin{equation}
        W_{i[a}{}^{[ke} \delta^f_b \delta^g_c \delta^{h]}_{d]} = 0,
\end{equation}
and then over $i$ and $k$ gives
\begin{equation}
        W_{[ab}{}^{[ef} \delta^g_c \delta^{h]}_{d]} = 0.
\end{equation}
A third contraction gives zero on the left hand side also.
\smallskip

So, in each of the three cases, the first identity is obvious in the
sense that it is an explicit antisymmetrisation over $n+1$ indices in $n$
dimensions. What
is particularly interesting, and at first sight perhaps surprising in these
situations, is the existence of simple identities in
$n$ dimensions which involve explicit antisymmetrization over only $n-1$
indices. But of course, in addition, there is antisymmetrisation on 
both lower and upper
indices and the comparatively simple versions are due to the vanishing of the
trace of $W$.
\end{example}

\begin{example}\label{ex:Torsion}
Consider the trace-free $(2,1)$-form $L_{ab}{}^c$.

When we apply the above arguments we obtain,

\smallskip
\underbar{In three dimensions:}
$L_{[fa}{}^g \delta_b^d \delta_{c]}^e = 0.$

Contracting once on  the upper index on $L$ gives
\begin{equation}
        L_{[ab}{}^{[d} \delta_{c]}^{e]} = 0,
\end{equation}
and contracting once more gives zero on the left hand side also.

\smallskip
\underbar{In four dimensions:}
$L_{[ha}{}^i \delta_b^e \delta_c^f \delta_{d]}^g = 0.$

Contracting once on  the upper index on $L$ gives
\begin{equation}
        L_{[ab}{}^{[e} \delta_c^f \delta_{d]}^{g]} = 0,
\end{equation}
and once more gives zero on the left hand side also.

\smallskip
So, once again, we obtain identities in $n$ dimensions involving explicit
antisymmetrization over less than $n+1$ indices; although in this case it
involves $n$ indices.
\end{example}

\begin{example}\label{ex:Maxwell}
Consider the trace-free $(1,1)$-form $S_{a}{}^b$.

When we apply the above
arguments we obtain,

\smallskip
\underbar{In three dimensions:}
$S_{[g}{}^h \delta_a^d \delta_b^e \delta_{c]}^f = 0.$

Contracting once on  the upper index on $S$ gives
\begin{equation}
        S_{[a}{}^{[d} \delta_b^e \delta_{c]}^{f]} = 0,
\end{equation}
and contracting once more gives zero on the left hand side also.

\smallskip
\underbar{In four dimensions:}
$S_{[i}{}^j \delta_a^e \delta_b^f \delta_c^g \delta_{d]}^h = 0.$

Contracting once on  the upper index on $S$ gives
\begin{equation}
        S_{[a}{}^{[e} \delta_b^f \delta_c^g \delta_{d]}^{h]} = 0,
\end{equation}
and once more gives zero on the left hand side also.

So, once again, we obtain identities in $n$ dimensions involving
explicit antisymmetrization over only $n$ indices.
\end{example}

In the above examples we have obtained obvious identities by
antisymmetrisation over $n+1$ indices in $n$ dimensions; but in
addition, as a result of taking traces, we have obtained less obvious
identities containing \emph{less} deltas and explicit symmetrisation
over \emph{less than} $n+1$ indices. The existence of these additional
identities, which contain less deltas than the obvious ones, are very
important for building up more complicated identities. For example,
consideration of the obvious four dimensional identity for the Weyl
tensor $C_{[ab}{}^{fg}\delta_c^h\delta_d^i\delta_{e]}^j=0$ (with $10$
free indices), would not suggest the possibility of any third order
\emph{scalar} identities for $C$; on the otherhand, the less obvious
one formed from the double trace, $C_{[ab}{}^{[fg}\delta_{c]}^{h]}=0$
(with $6$ free indices), certainly permits the constructions of third
order Weyl scalar identities by multiplication with a pair of $C$
tensors.

For each case considered we have gone as far as we could, in the sense
that taking an additional trace simply reduces the left hand side to
zero. Nevertheless, this does not in itself mean that there could not
exist additional identities with even less deltas than those given
above; however, for each of the cases it can be shown directly that
removing further deltas gives a restriction rather than an identity,
e.g., \emph{in four dimensions}, when the trace is calculated for
$L_{[ab}{}^{[c} \delta_{f]}^{g]} = 0$ we find that $L_{ab}{}^{c}=0$.

In the next section we shall give results due to
Lovelock~\cite{lovelock70} and show that the results for three and
four dimensions of Example~\ref{ex:Weyl} follow directly as a special
case; it will also be shown that the result for five dimensions can be
deduced, more subtly, from Lovelock's results.  In
Section~\ref{ch:basth} we will derive even more general results of
which Lovelock's and the above three examples will be seen to be
special cases.

\section{Lovelock's Dimensionally Dependent Identities }\label{ch:ldimdepid}

Some years ago, Lovelock~\cite{lovelock70} noted that familiar
tensors, (such as the Weyl and Maxwell tensors), with antisymmetry and
trace-free properties, obeyed assorted --- apparently unrelated ---
identities.  However, although there was no common structural link in
the original derivations of these assorted identities,
Lovelock~\cite{lovelock70} showed that they could
all be considered to be consequences of two  underlying basic tensor
identities. These underlying identities had a very simple structure
and were a mathematically trivial consequence of dimension alone. We 
will now quote
Lovelock's two theorems, each of which he proved in two different 
ways; our proofs are
essentially conciser presentations of  one of his versions.

\begin{theorem}\label{th:lovelock1}
In an $n$-dimensional space let $T^{\cal{A}}{}_{a_1\ldots 
a_k}{}^{b_1\ldots b_k} =
T^{\cal{A}}{}_{[a_1\ldots a_k]}{}^{[b_1\ldots b_k]}$ be trace-free on 
its explicit
indices.  If
$ 2k > n$ then
\begin{equation}
        T^{\cal{A}}{}_{a_1\ldots a_k}{}^{b_1\ldots b_k}=0
\end{equation}
\end{theorem}
\begin{proof}
Since $2k > n$, antisymmetrising over $2k$ indices gives an
identity
\begin{equation}
        0 =
        T^{\cal{A}}{}_{[a_1\ldots a_k}{}^{i_1\ldots i_k}
        \delta_{i_1}^{b_1}\ldots\delta_{i_k]}^{b_k}.
\end{equation}
Since the tensor $T$ is trace-free, we get
\begin{equation}
        0 =
        T^{\cal{A}}{}_{a_1\ldots a_k}{}^{i_1\ldots i_k}
        \delta_{i_1}^{b_1}\ldots\delta_{i_k}^{b_k}.
\end{equation}
Absorbing the deltas gives the theorem.
\end{proof}

\begin{theorem}\label{th:lovelock2}
In an $n$-dimensional space let $T^{\cal{A}}{}_{a_1\ldots 
a_k}{}^{b_1\ldots b_k} =
T^{\cal{A}}{}_{[a_1\ldots a_k]}{}^{[b_1\ldots b_k]}$ be trace-free on 
its explicit
indices.  If $2k=n$ then
\begin{equation}\label{eq:lovelock2}
        T^{\cal{A}}{}_{[a_1\ldots a_k}{}^{[b_1\ldots
b_k}\delta_{a_{k+1}]}^{b_{k+1}]}=0
\end{equation}
\end{theorem}
\begin{proof}
The proof is analogous to the proof of Theorem~\ref{th:lovelock1}
but this time the antisymmetrisation is over $2k+1$ indices.  Starting
with
\begin{equation}
        0 =
        T^{\cal{A}}{}_{[a_1\ldots a_k}{}^{i_1\ldots i_k}
        \delta_{a_{k+1}}^{b_{k+1}}\delta_{i_1}^{b_1}\ldots\delta_{i_k]}^{b_k}
        =
        T^{\cal{A}}{}_{[a_1\ldots a_k}{}^{i_1\ldots i_k}
        \delta_{a_{k+1}}^{[b_{k+1}}\delta_{i_1}^{b_1}\ldots\delta_{i_k]}^{b_k]},
\end{equation}
since the tensor is trace-free, we get
\begin{equation}
        0 =
        T^{\cal{A}}{}_{[a_1\ldots a_k}{}^{i_1\ldots i_k}
        \delta_{a_{k+1}]}^{[b_{k+1}}\delta_{i_1}^{b_1}\ldots\delta_{i_k}^{b_k]}.
\end{equation}
Absorbing the deltas gives the theorem.
\end{proof}

   (This proof  shows that this second theorem is actually true for 
the weaker condition,
$2k\geq n$; but the validity of the second theorem when $2k > n$ also 
follows from the
first theorem, which has the more fundamental condition.)

The theorems immediately yield the familiar results in
Example~\ref{ex:Weyl}:
\begin{example}\label{ex:Lovelock}
Theorem~\ref{th:lovelock1} with $k=2$ applied to the trace-free
$(2,2)$-form $W_{abcd}$ gives $W_{abcd}=0$ when $n\leq 3$.

From Theorem~\ref{th:lovelock2} we find directly $
W^{[cd}{}_{[ef}\delta_{b]}^{a]}=0$ when $n\leq 4$.

An additional well-known result is found by multiplying this
with $W^{ef}{}_{cd}$ to get
\begin{equation}\label{eq:Weylsquare}
        W^{ac}{}_{de}W^{de}{}_{bc}
        =
        \frac{1}{4}\delta_b^a W^{cd}{}_{ef}W^{ef}{}_{cd}.
\end{equation}
Multiplying with $W^{ef}{}_{ga} W^{bg}{}_{cd}$ instead yields the scalar
identity cubic in $W^{ab}{}_{cd}$,
\begin{equation}\label{eq:Weylcubic}
        W^{ab}{}_{ce} W^{cd}{}_{af} W^{ef}{}_{bd}
        =
        \frac{1}{4} W^{ab}{}_{cd} W^{cd}{}_{ef} W^{ef}{}_{ab}.
\end{equation}

The third result in Example \ref{ex:Weyl} may also be obtained ---
but in a more indirect manner; we consider the tensor
$T^{abef}{}_{cdkl} = W^{[ab}{}_{[cd}\delta^e_k \delta^{f]}_{l]}$ and
\emph{by a direct calculation} we can confirm its trace to be zero in five
dimensions. Hence, from Theorem~\ref{th:lovelock1}, it follows
immediately that $T^{abef}{}_{cdkl}$ is identically zero in five
dimensions.

The results in Example~\ref{ex:Maxwell} can also be obtained in the
same way. 
\end{example}

As well as showing how familiar identities were direct consequences of
his theorems, Lovelock~\cite{lovelock70} also deduced some new
interesting identities. More recently, additional important, and
sometimes unexpected, identities have been shown to follow from
Lovelock's theorems,~\cite{edgar94,ae,edgar99,eh00}.

However, we emphasise again that the results in
Example~\ref{ex:Torsion} and~\ref{ex:Maxwell} cannot be deduced
\emph{directly} from Lovelock's theorems.

\section{General theorems on dimensionally dependent
identities}\label{ch:basth}

The natural generalisation of Lovelock's theorems is the following theorem.

\begin{theorem}\label{th:trfree2}
In an $n$-dimensional space let $T^{\cal{A}}{}_{a_1\ldots
a_k}{}^{b_1\ldots b_l} = T^{\cal{A}}{}_{[a_1\ldots a_k]}{}^{[b_1\ldots b_l]}$
be trace-free on its explicit indices.
Then
\begin{equation}\label{eq:gen1}
          T^{\cal{A}}{}_{[a_1\ldots a_k}{}^{[b_1\ldots b_l}
          \delta_{a_{k+1}}^{b_{l+1}}\ldots\delta_{a_{k+d}]}^{b_{l+d}]} = 0
\end{equation}
where $d \geq n-k-l+1$ and $d \geq 0$.
\end{theorem}

\begin{proof}
The proof is analogous to the proof of
Theorem~\ref{th:lovelock2}.  This time we antisymmetrise over
$k+l+d\geq n+1$ indices to get the identity.
\begin{equation}
        0 =
        T^{\cal{A}}{}_{[a_1\ldots a_k}{}^{i_1\ldots i_l}
        \delta_{a_{k+1}}^{b_{l+1}}
        \ldots
        \delta_{a_{k+d}}^{b_{l+d}}
        \delta_{i_1}^{b_1}\ldots\delta_{i_l]}^{b_l}
        =
        T^{\cal{A}}{}_{[a_1\ldots a_k}{}^{i_1\ldots i_l}
        \delta_{a_{k+1}}^{[b_{l+1}}
        \ldots
        \delta_{a_{k+d}}^{b_{l+d}}
        \delta_{i_1}^{b_1}\ldots\delta_{i_l]}^{b_l]}.
\end{equation}
Since the tensor $T$ is trace-free, we get
\begin{equation}
        0 =
        T^{\cal{A}}{}_{[a_1\ldots a_k}{}^{i_1\ldots i_l}
        \delta_{a_{k+1}}^{[b_{l+1}}
        \ldots
        \delta_{a_{k+d}]}^{b_{l+d}}
        \delta_{i_1}^{b_1}\ldots\delta_{i_l}^{b_l]}.
\end{equation}
Absorbing the deltas gives the result.
\end{proof}

\noindent
\textbf{Remarks.}

\noindent
$\bullet$ For completeness, we add, that if we adopt the convention
that tensors with only lower or only upper indices are considered
'trace-free', then the above results also hold for the two classes of
tensors where $k= 0$ or $l=0$. In addition, for the case $k= 0$ and
$l=0$, i.e., $T$ a scalar, we simply get the trivial result of the
vanishing of the Kronecker delta symbol with $n+1$ index pairs in
$n$-dimensional space.

\noindent
$\bullet$ The theorem generalises Lovelock's results by associating,
in $n$-dimensional space, an identity with \emph{any} tensor, since
--- no matter what its index configuration --- part of it can be
considered as a $(k,l)$-form, with the above conventions when $k=0$ or
$l=0$. But it is for trace-free double forms with $k\ne 0 \ne l$ that
it will be most useful. So when $l=0$ or $k=0$ and $d=0$ we have
identity~\eqref{eq:id2}; when $k=l\ne 0$ and $d=0$ this is
Theorem~\ref{th:lovelock1} and when $k=l\ne 0$ and $d=1$ this is
Theorem~\ref{th:lovelock2}. Had we wished only to generalise to the
case $k=l$ for any $d$, we could have used Theorem~\ref{th:lovelock1}
directly in our proof, as in Example \ref{ex:Weyl}.

\noindent
$\bullet$ No stronger result on $T$ can be obtained by taking the
trace of~\eqref{eq:gen1}, since then the left hand side just collapses
to zero.  In addition, the discussion at the end of
Section~\ref{ch:dimdepid} would strongly suggest that the conditions
on $d$ can not be relaxed for non-zero $T$; we shall confirm later in
Theorem~\ref{th:inverse1} that this is indeed so.

\noindent
$\bullet$ Of course this theorem does not mean that we cannot
construct results beginning with double forms which are not
trace-free.  Rather, what will happen is that if we begin with such a
tensor we will get an apparently more complicated identity with
explicit trace terms; if the double form is then decomposed into
trace-free and trace parts, the resulting simplification will leave us
with the identity which would be obtained by beginning with the
trace-free part of the double form.

\

We shall now illustrate the relevance of the new results
in Theorem~\ref{th:trfree2} to familiar tensors as follows,

\begin{example}\label{ex:lanczos2}
Applying Theorem~\ref{th:trfree2} to the  trace-free
$(2,1)$-form $L_{ab}{}^c$ ($l=1, k=2$) in
dimensions
$n=3$ with $d=1$ gives the identity
\begin{equation}
        L_{[ab}{}^{[d}\delta_{c]}^{e]}=0
\end{equation}
and in dimensions
$n=4$,  with $d=2$
\begin{equation}
        L_{[ab}{}^{[e}\delta_c^f\delta_{d]}^{g]}=0.
\end{equation}
For  trace-free $(2,2)$-forms
($l=k=2$)
in  dimensions
$n=5$ with
$d=2$ gives the identity
\begin{equation}\label{eq:WeyI3}
   W_{[ab}{}^{[ef}\delta_c^g\delta^{h]}_{d]} = 0.
\end{equation}
For trace-free $(1,1)$-forms
$S^a{}_b$  ($l=k=1$), in  dimensions
$n=4$ with
$d=3$ gives the identity
\begin{equation}\label{eq:Ricci1}
   S_{[a}{}^{[e}\delta^f_b\delta^g_c\delta^{h]}_{d]} = 0.
\end{equation}
\end{example}
Analogous basic identities can be found in other dimensions, and all
such identities can then be exploited to build up other useful
important identities, as we shall demonstrate in the next section.

We now consider whether we can generalise these results in another
manner, by asking whether any kind of converses exist.  Firstly, by
way of example, we consider the identity~\eqref{eq:WeyI3} obtained
from Theorem~\ref{th:trfree2} for the trace-free $(2,2)$-forms when
$n\leq 5$. As mentioned in the third
remark above, we would like to confirm explicitly that the stronger
condition $W_{[ab}{}^{[de} \delta_{c]}^{f]}=0$, which holds in four
dimensions, does not hold in dimensions $n=5$.  A related task would
be to confirm explicitly that the same identity~\eqref{eq:WeyI3} as
holds in $n\leq 5$ dimensions does not hold in higher dimensions.  We
shall see from Theorem \ref{th:inverse1} that we can confirm
explicitly an even stronger version of these results.

Secondly, with respect to the same example, a natural question to ask
is whether the trace-free condition is also a necessary condition,
i.e., whether any forms with \emph{non-zero trace} can satisfy this
identity $ W_{[ab}{}^{[ef}\delta_c^g\delta_{d]}^{h]}=0$ in
dimensions $n\leq 5$. Whether an alternative $(2,2)$-form --- lacking
the trace-free properties but perhaps with different symmetry
properties --- can satisfy the basic identity is not obviously ruled
out. As an example, we could ask whether a $(2,2)$-form like the
Riemann tensor $R_{ab}{}^{cd}$ --- lacking the trace-free properties
of $W_{ab}{}^{cd}$, but having the additional properties
$R_{a[bcd]}=0$ and $R_{abcd}=R_{cdab}$ --- can satisy the
identities. However, we shall see from Theorem~\ref{th:inverse2} that
\emph{only trace-free} $(2,2)$-forms satisfy the
identity~\eqref{eq:WeyI3} in dimensions $n\leq 5$.

We first present a lemma which is then used in the proof of the two
theorems. We emphasise that this lemma is also useful in its own
right, and
we shall demonstrate how it can be viewed as a generalisation of a familiar
result for the Kronecker delta.

\begin{lemma}\label{lemma:trace1}
In $n$-dimensional space let $T^{\cal{A}}{}_{a_1\ldots
a_k}{}^{b_1\ldots b_l}=T^{\cal{A}}{}_{[a_1\ldots a_k]}{}^{[b_1\ldots
b_l]}$, $k$, $l$ and $d\geq 0$ and $(k+d)(l+d)>0$.  Then
\begin{equation}
\begin{split}
        T^{\cal{A}}{}_{[a_1\ldots a_k}{}^{[b_1\ldots b_l}
        \delta_{a_{k+1}}^{b_{l+1}}
        \ldots\delta_{a_{k+d}]}^{b_{l+d}]}
        \delta^{a_{k+d}}_{b_{l+d}}
        = &
        \frac{d(n-(d+k+l-1))}{(k+d)(l+d)}
        T^{\cal{A}}{}_{[a_1\ldots a_k}{}^{[b_1\ldots b_l}
        \delta_{a_{k+1}}^{b_{l+1}}\ldots
        \delta_{a_{k+d-1}]}^{b_{l+d-1}]}
        \\ &
        +\frac{(-1)^{k+l} kl}{(k+d)(l+d)}
        T^{\cal{A}}{}_{c[a_1\ldots a_{k-1}}{}^{c[b_1\ldots b_{l-1}}
        \delta_{a_{k}}^{b_{l}}\ldots
        \delta_{a_{k+d-1}]}^{b_{l+d-1}]}.
\end{split}
\end{equation}
\end{lemma}
\begin{proof}
If $k$, $l$ or $d$ is zero the lemma is trivial.  Assume they are
non-zero.

On the left hand side the last two deltas combine to give
\begin{equation}
        T^{\cal{A}}{}_{[a_1\ldots a_k}{}^{[b_1\ldots b_l}
        \delta_{a_{k+1}}^{b_{l+1}}
        \ldots
        \delta_{a_{k+d-1}}^{b_{l+d-1}}
        \delta_{c]}^{c]}.
\end{equation}

By summing over all possible positions of the dummy indices $c$ we
get
\begin{equation}
        \frac{(k+d-1)!(l+d-1)!}{(k+d)!(l+d)!}
        \sum_{i,j} c_{ij}
        T^{\cal{A}}{}_{[a_1\ldots a_k}{}^{[b_1\ldots b_l}
        \delta_{a_{k+1}}^{b_{l+1}}
        \ldots
        \delta_{|c|}^{b_{k_i}}
        \ldots
        \delta_{a_{l_j}}^{|c|}
        \ldots
        \delta_{a_{k+d-1}]}^{b_{l+d-1}]}
\end{equation}
where $c_{ij}=\pm1$ depending on whether the index configuration is an
even or an odd permutation.

It is now clear that we get only the two types of terms that are on
the right hand side of the theorem; the last type when both dummy
indices are on $T$ and the other type when at least one of them is on
a $\delta$.  What remains is to confirm the coefficients, which
essentially means counting how many of each kind occurs and what sign
they have.

When both dummy indices are on $T$ they can be moved to the first
position.  That means that they have moved a total of
$(k+d-1)+(l+d-1)$ positions and changed sign equally many times.  That
gives a factor $(-1)^{k+l}$.  There are a total of $kl$ such terms.

When both dummy indices are on the same $\delta$ they have moved equally
many positions; thus such terms are added.  There are $d$ such
terms and we also get a factor $n$ since $\delta_i^i=n$.

When both dummy indices are on different $\delta$ one of them can be
absorbed thereby distorting the order of the other indices.  Once that
order is restored we have overall an odd permutation; thus such terms
are subtracted.  There is a total of $d(d-1)$ such terms.

The same situation occurs when one of the dummy indices is on a $\delta$
and the other one is on $T$.  There are $ld$ such terms with the upper
index on $T$ and $kd$ with the lower index on $T$.

Taking all this together gives the identity in the lemma.
\end{proof}
When we consider  $k=0$ and $l=0$, with $T$  a
non-zero
constant, in
Lemma \ref{lemma:trace1} we get a familiar identity for the Kronecker delta,

\begin{corollary}\label{th:delta}
In $n$-dimensional space
\begin{equation}
        \delta_{[a_{1}}^{[b_{1}}
        \ldots\delta_{a_{d}]}^{b_{d}]}
        \delta^{a_{d}}_{b_{d}}
        =
        \frac{(n-d+1)}{d}
        \delta_{[a_{1}}^{[b_{1}}\ldots
        \delta_{a_{d-1}]}^{b_{d-1}]}.
\end{equation}
\end{corollary}

We now use Lemma~\ref{lemma:trace1} to establish the two theorems.

\begin{theorem}\label{th:inverse1}
In an $n$-dimensional space let  $T^{\cal{A}}{}_{a_1\ldots 
a_k}{}^{b_1\ldots b_l}=
T^{\cal{A}}{}_{[a_1\ldots a_k]}{}^{[b_1\ldots b_l]}$.  If
\begin{equation}\label{eq:bignth}
        T^{\cal{A}}{}_{[a_1\ldots a_k}{}^{[b_1\ldots b_l}
        \delta_{a_{k+1}}^{b_{l+1}}\ldots\delta_{a_{k+d}]}^{b_{l+d}]} = 0
\end{equation}
then
\begin{equation}
        n >  d+k+l-1 \implies
        T^{\cal{A}}{}_{a_1\ldots a_k}{}^{b_1\ldots b_l} = 0.
\end{equation}
\end{theorem}
\begin{proof}
The basic idea is to repeatedly take traces of
equation~\eqref{eq:bignth} getting a sequence of equations; in the last
of these no $\delta$ remains.  The result follows from substituting
equations at the end of this sequence into earlier ones.  However,
care must be taken so that no unwanted cancelling of terms occur which
forces this process to stop prematurely.

To make the proof easier to overview we define the following notation.
Let $(d,k,l) = T^{\cal{A}}{}_{[a_1\ldots a_k}{}^{[b_1\ldots b_l}
\delta_{a_{k+1}}^{b_{l+1}}\ldots\delta_{a_{k+d}]}^{b_{l+d}]}$ and if
$k$ and $l$ are less then the actual number of indices on $T$ then we
contract over the remaining ones.  Thus Lemma~\ref{lemma:trace1} can
be written as $\text{the trace of }(d,k,l)=
\frac{d(n-(d+k+l-1))}{(k+d)(l+d)}(d-1,k,l) +\frac{(-1)^{k+l}
kl}{(k+d)(l+d)}(d,k-1,l-1)$ and $T$ itself as $(0,k,l)$ and the trace
of $T$ as $(0,k-1,l-1)$.

We observe that the coefficients in Lemma~\ref{lemma:trace1} are
non-zero if $d$ is non-zero and $k$ and $l$ are non-zero respectively.

First assume that $k+l$ is even, then the coefficients in
Lemma~\ref{lemma:trace1} are non-negative.  Start with
equation~\eqref{eq:bignth} and multiply with
$\delta_{b_{l+d}}^{a_{k+d}}$ and use Lemma~\ref{lemma:trace1}.  Then we
get $0=(d-1,k,l)+(d,k-1,l-1)$ where we have omitted the coefficients.

Doing that once more gives us $0=(d-2,k,l)+(d-1,k-1,l-1)+(d,k-2,l-2)$.
Repeating this process gives a sequence of equations as illustrated
in Figure~\ref{fig:id} where each row corresponds to one equation.

\begin{figure}[h]
\begin{center}
{\footnotesize
\begin{picture}(95,80)
\leftlegtrue
\rightlegtrue

\put(25,65){\legs{$(d,k,l)$}}

\leftdottrue
\put(10,50){\legs{$(d-1,k,l)$}}
\leftdotfalse
\put(40,50){\legs{$(d,k-1,l-1)$}}

\leftlegfalse
\put(-5,35){\legs{$(0,k,l)$}}
\leftlegtrue
\leftdottrue
\put(25,35){\legs{$(d-1,k-1,l-1)$}}
\leftdotfalse
\rightdottrue
\put(55,35){\legs{$(d,k-2,l-2)$}}

\leftlegfalse
\put(10,20){\legs{$(0,k-1,l-1)$}}
\leftlegtrue
\leftdottrue
\put(40,20){\legs{$(d-1,k-2,l-2)$}}
\rightlegfalse
\put(70,20){\legs{$(d,k-l,0)$}}
\rightlegtrue
\leftdotfalse
\rightdotfalse

\leftlegfalse
\put(25,5){\legs{$(0,k-l+1,1)$}}
\leftlegtrue
\rightlegfalse
\put(55,5){\legs{$(1,k-l,0)$}}

\leftlegfalse
\put(40,-10){\legs{($0,k-l,0)$}}
\end{picture}
}
\end{center}
\caption{Sequence of equations}
\label{fig:id}
\end{figure}
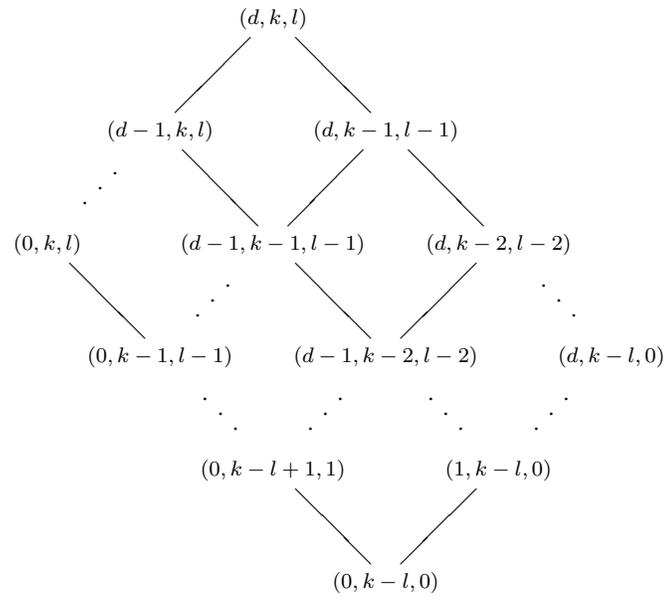

The last equation is $0=(0,k-l,0)$ (assuming $k\geq l$).  Putting this
into the second last equation gives $0=(0,k-l+1,1)$ by using that
$0=(j,k,l) \implies 0=(j+1,k,l)$.  Feeding the new information into
earlier equations gives the desired conclusion $0=(0,k,l)$.

If $k+l$ is odd then there is a minus sign in Lemma~\ref{lemma:trace1}
which means that there is a risk of cancellation in the above process.
However, checking how the signs propagate gives us
Figure~\ref{fig:sign} where the sign at a node is the sign of the term
and the sign at the edge is the sign of the coefficient in the
identity in Lemma~\ref{lemma:trace1}.

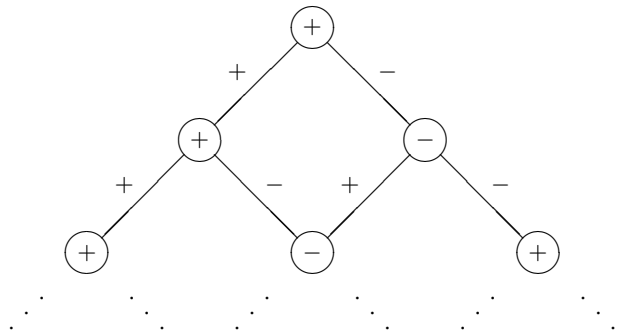
\begin{figure}[h]
\begin{center}
\begin{picture}(90,47)
\leftlegtrue
\leftdotfalse
\rightlegtrue
\rightdotfalse
\signtrue

\put(30,30){\legs{$+$}}

\put(15,15){\legs{$+$}}
\put(45,15){\legs{$-$}}

\rightdottrue
\leftdottrue
\put(0,0){\legs{$+$}}
\put(30,0){\legs{$-$}}
\put(60,0){\legs{$+$}}
\end{picture}
\end{center}
\caption{Signs}
\label{fig:sign}
\end{figure}

It is now clear that there will be no cancellations since each term
originates as the difference between two terms with different sign.
\end{proof}

\begin{theorem}\label{th:inverse2}
In an $n$-dimensional space let $T^{\cal{A}}{}_{a_1\ldots 
a_k}{}^{b_1\ldots b_l} = T^{\cal{A}}{}_{[a_1\ldots a_k]}{}^{[b_1\ldots
b_l]}$ where $n\geq k+l-1$ and let $d=n-k-l+1$.  Then
\begin{equation}
   T^{\cal{A}}{}_{[a_1\ldots a_k}{}^{[b_1\ldots b_l}
          \delta_{a_{k+1}}^{b_{l+1}}\ldots\delta_{a_{k+d}]}^{b_{l+d}]}=0
        \iff
        T^{\cal{A}}{}_{a_1\ldots a_k}{}^{b_1\ldots b_l} \text{ is
trace-free on its explicit
indices}.
\end{equation}
\end{theorem}

\begin{proof}
$\impliedby$:  Follows directly from Theorem~\ref{th:trfree2}

$\implies$:
The case $d=0$ is trivial.  Assume $d>0$.  Contracting
\begin{equation}
          0 =
          T^{\cal{A}}{}_{[a_1\ldots a_k}{}^{[b_1\ldots b_l}
          \delta_{a_{k+1}}^{b_{l+1}}\ldots\delta_{a_{k+d}]}^{b_{l+d}]}
\end{equation}
once and using Lemma~\ref{lemma:trace1} gives
\begin{equation}
        0 =
        T^{\cal{A}}{}_{i[a_1\ldots a_{k-1}}{}^{i[b_1\ldots b_{l-1}}
        \delta_{a_{k}}^{b_{l}}\ldots
        \delta_{a_{k+d-1}]}^{b_{l+d-1}]}.
\end{equation}
The theorem now follows from Theorem~\ref{th:inverse1} applied to the
tensor
\begin{equation}
        \tilde{T}_{a_1\ldots a_{k-1}}{}^{b_1\ldots b_{l-1}}
        =
        T^{\cal{A}}{}_{ia_1\ldots a_{k-1}}{}^{ib_1\ldots b_{l-1}}.
\end{equation}
\end{proof}

\begin{example} We know that for the Weyl curvature tensor $C_{abcd}$ 
(a trace-free symmetric $(2,2)$-form), $
C_{[ab}{}^{[ef}\delta_c^g\delta_{d]}^{h]}=0$ in dimensions $n\leq
5$. From Theorem~\ref{th:inverse1} we can conclude that this tensor
(or indeed \emph{any} non-zero $(2,2)$-form) cannot satisfy the
stronger condition $ C_{[ab}{}^{[de}\delta_{c]}^{f]}=0$ in dimensions
$n\geq 5$; in addition we can conclude that there are absolutely no
non-zero $(2,2)$-forms $T_{ab}{}^{cd}$ satisfying
$T_{[ab}{}^{[ef}\delta_c^g\delta_{d]}^{h]}=0$ \emph{in dimensions
greater than five}.

From Theorem~\ref{th:inverse2} we can conclude that there are no
$(2,2)$-forms $R_{ab}{}^{cd}$ \emph{with non-zero trace} (e.g., a
Riemann tensor) satisfying the identity
$R_{[ab}{}^{[ef}\delta_c^g\delta_{d]}^{h]}=0$ in dimensions $n = 5$.
\end{example}

The trace in Lemma~\ref{lemma:trace1} is not the only trace that is 
possible for
expressions of the type we are investigating.  For completeness we here
present the other possibility.

\begin{lemma}
Let $T^{\cal{A}}{}_{a_1\ldots a_k}{}^{cb_1\ldots b_l} =
T^{\cal{A}}{}_{[a_1\ldots a_k]}{}^{c[b_1\ldots b_l]}$, $k$ and $d\geq
0$ and $k+d>0$.  Then
\begin{equation}\label{eq:trace2}
\begin{split}
        T^{\cal{A}}{}_{[a_1\ldots a_k}{}^{c[b_1\ldots b_l}
                \delta_{a_{k+1}}^{b_{l+1}} \ldots
                \delta_{a_{k+d}]}^{b_{l+d}]} \delta_c^{a_{k+d}}
         = &
        \frac{d(-1)^{l+d-1}}{k+d}
                T^{\cal{A}}{}_{[a_1\ldots a_k}{}^{[b_1\ldots b_lb_{l+1}}
                \delta_{a_{k+1}}^{b_{l+2}} \ldots
                \delta_{a_{k+d-1}]}^{b_{l+d}]}
        \\ &
        +\frac{k(-1)^{k+d-1}}{k+d}
                T^{\cal{A}}{}_{c[a_1\ldots a_{k-1}}{}^{c[b_1\ldots b_l}
                \delta_{a_k}^{b_{l+1}} \ldots
                \delta_{a_{k+d-1}]}^{b_{l+d}]}.
\end{split}
\end{equation}
\end{lemma}
\begin{proof}
If $k$ or $d$ is zero the lemma is trivial.  Assume they are non-zero.

Absorbing the last delta on the left hand side gives
\begin{equation}
        T^{\cal{A}}{}_{[a_1\ldots a_k}{}^{c[b_1\ldots b_l}
                \delta_{a_{k+1}}^{b_{l+1}} \ldots
                \delta_{c]}^{b_{l+d}]}.
\end{equation}
By summing over all possible positions of the lower dummy index $c$ we
get
\begin{equation}
        \frac{(k+d-1)!}{(k+d)!} \sum_i c_i
        T^{\cal{A}}{}_{[a_1\ldots a_k}{}^{c[b_1\ldots b_l}
        \delta_{a_{k+1}}^{b_{l+1}} \ldots \delta_{|c|}^{b_i} \ldots
        \delta_{a_{k+d-1}]}^{b_{l+d}]}
\end{equation}
where $c_i=\pm 1$ depending on whether the index configuration is an
even or an odd permutaion.

When the $c$ is on a delta it can be absorbed giving terms of the same
type as the first term on the right hand side of~\eqref{eq:trace2}.
There are $d$ such terms and on each of them the indices have been
moved a total of $l+d-1$ steps.

When the $c$ is on $T$ it can be moved to the first position of the
lower indices giving terms of the same type as the second term on the
right hand side of~\eqref{eq:trace2}.  There are $k$ such terms and on
each of them the index has been moved a total of  $k+d-1$ steps.
\end{proof}

\section{Applications.}\label{ch:abi}

In the last sections we have noted that, associated with each tensor, 
are a number of
fundamental identities; and it is from such identities that more involved
and more subtle identities can be constructed, which in turn yield 
familiar identities.
So, we shall now exploit our results  in two particular types of applications:

\subsection{Identities involving scalar invariants of Riemann tensors.}

Relationships between scalar invariants of tensors (such as
\eqref{eq:xu1}, \eqref{eq:xu2} and~\eqref{eq:Weylcubic}) play a very
important role in classical invariant theory, as well as in many
practical applications.  For instance, the study of the scalar
invariants of the Riemann tensor in four dimensions has posed many
interesting problems, which have not all been
resolved~\cite{F,bonanos98,harvey95,jack87,ouchterlony97,sneddon96,sneddon98,zakhary97}.
The theorems given in this paper provide important tools for a
systematic study of invariants of the Weyl, trace-free Ricci and
Lanczos tensors. However, now we shall just apply our results to some
representative examples.

\begin{example}
Although we can form different scalars from two $(2,2)$-forms
$W_{ab}{}^{cd}$, in \emph{four dimensions}, the simplest basic
identity is $ W^{[cd}{}_{[ef}\delta_{b]}^{a]}=0$ (with $6$ free
indices), and so we can immediately see that it cannot yield
relationships between scalar invariants involving only two
$(2,2)$-forms.

However, we already know that multiplying this identity by
$W^{ef}{}_{ga} W^{bg}{}_{cd}$ gives us the scalar identity
(\ref{eq:Weylcubic}), which yields a relationship between some cubic
scalar invariants; while multiplying by other quadratic terms will
give different relationships between different cubic scalar
invariants. Clearly we can also choose various suitable expressions
involving three $(2,2)$-forms, which will yield scalar identities of
fourth order when multiplied with $W^{[cd}{}_{[ef}\delta_{b]}^{a]}=0$.
Hence, it is possible to investigate, for each order, all such
possible relationships between all scalar invariants of that order.

When we consider higher dimensions we have analogous basic
identities. In \emph{five dimensions} the simplest basic identity is $
W_{[ab}{}^{[ef}\delta_c^g\delta_{d]}^{h]}=0$ (with eight free
indices), and hence the lowest order where we can get relationships
between scalar invariants from this identity is also at third order;
while in \emph{six dimensions} with the simplest basic identity having
$10$ free indices there can exist no relationships between scalar
invariants at third order coming from this identity.  These results
can be applied to the Weyl tensor, where they may simplify a little
because of its extra symmetries.
\end{example}

It is interesting to note that Dianyan Xu's work~\cite{xu87} was
motivated by a concern that non-trivial relationships might exist
between counterterms in the Lagrangian (essentially invariant scalars
constructed from products of the Riemann tensor) of a renormalizable
quantum field theory; in particular, if his identities~\eqref{eq:xu1},
\eqref{eq:xu2}, or any other cubic scalar identities, are true in six
dimensions then work by Jack and Parker~\cite{jack85} would need to be
revaluated. However, Jack and Parker~\cite{jack87} have subsequently
shown explicitly that no such third order identities can exist in six
dimensions; our result in the example above agrees with this.  Jack
and Parker~\cite{jack87} have conjectured that in $2n$ dimensions
there do not exist any identities between Riemann scalars of order
$n$; we shall show how our results relate to this conjecture in a
subsequent paper.

\begin{example}
In order to confirm that the second identity found in~\cite{xu87} was
a four dimensional one, Harvey first established the intermediate
third order identity~\eqref{eq:h2} by antisymmetrising over \emph{six}
indices and this intermediate identity is therefore valid in five as
well as in four dimensions; in our previous example we noted that five
dimensions was the lowest dimension where such third order identities
could be constructed for the Weyl tensor.  However, one would suspect
that the second four dimensional identity~\eqref{eq:xu2} could be
obtained directly; this is confirmed by using the four dimensional
identity $C^{[cd}{}_{[ef}\delta_{b]}^{a]}=0$ and expanding
$C^{ef}{}_{ai} C^{ib}{}_{cd} C^{[cd}{}_{[ef}\delta_{b]}^{a]} = 0$.  It
can also be obtained, in the manner of Harvey, by antisymmetrising
over 5 indices and expanding
\begin{equation}
        \Ruull ab{[a}b \Ruull efcd \Ruull cd{e]}f
        =0.
\end{equation}
\end{example}

\begin{example}
An important application of the identity in Example~\ref{ex:lanczos2}
is to find relationships between scalar invariants of the Lanczos
potential~\cite{lanczos62} in four dimensions.  Since the identity
$L_{[ab}{}^{[e} \delta_c^f\delta_{d]}^{g]} = 0$ has seven free
indices, the lowest order relationship between scalar invariants that
we can obtain from it is of order four.  For example, by expanding
\begin{equation}
        0=
        L_{[ab}{}^{[e} \delta_c^f\delta_{d]}^{g]}
        L^{ab}{}_e
        L_{fg}{}^h
        L^{cd}{}_h ,
\end{equation}
we obtain the relationship
\begin{equation}
        0=
        L_{abc}L^{abc}L_{def}L^{def}
        +L_{abc}L^{abf}L_{def}L^{dec}
        -4L_{afc}L^{aec}L_{deh}L^{dfh}
        -2L_{abc}L^{abd}L_{def}L^{cef}
        -4L_{abc}L^{ad}{}_fL^{bfh}L^c{}_{dh} .
\end{equation}
This of course will not be the only relationship; we could instead multiply the
original identity by
$ L^{ab}{}_e L_{fh}{}^c L_g{}^{hd}$.
Therefore,  we can investigate all possible
contractions of
the original identity with three
Lanczos tensors and find the corresponding
relationship between all possible quartic Lanczos scalars in a 
systematic manner.
\end{example}

There is an important caveat in the above examples. By a systematic
study we are able to obtain all relationships between scalar
invariants of a particular order \emph{which arise from our
dimensionally dependent identities.}  Of course, we would like to be
able to conclude that we have obtained \emph{all possible} such
relationships.  Recent work on invariant theory by
Gover~\cite{gover96} links relationships involving scalar invariants
of tensors in $n$ dimensions to antisymmetrising over $n+1$
indices. Although this does not, at this stage, enable us to conclude
that all relationships between scalar invariants originate from our
dimensionally dependent identities, it would lead us to believe that
the results in this paper will be useful in such difficult tasks as
determining complete and independent sets of scalar invariants, and
their syzygies.

\subsection{Simplifying complicated expressions in four  dimensions.}

There have recently been situations~\cite{ae,edgar94,edgar99,eh00}
where rather complicated tensor expressions in $n$ dimensions have
been shown, unexpectedly, to be identically zero when specialised to
four dimensions, via Lovelock's theorems. So we would anticipate that
the more general results obtained in this paper should also be useful
in simplifying such expressions, and we now give some examples.

\begin{example}\label{ex:sen2}
In~\cite{ae} Lovelock's identities were used to show that, when the
wave equation for the Weyl tensor is constructed from the Bianchi
identities, the sum of terms which involve products of Weyl and
trace-free Ricci tensors disappeared in \emph{four, and only four,
dimensions}, because from the $4$-dimensional identity
$C_{[ab}{}^{[de}\delta_{c]}^{f]}=0$ we can obtain
$C_{[ab}{}^{[de}\delta_{c]}^{f]}\tilde R_f{}^c=0$ whose left hand
side, when expanded, is precisely this sum of terms.

We shall now show that the analogous component of the Bel
tensor~\cite{Bel} disappears in four dimensions by virtue of the same
identity.  The Bel tensor is given in $n$ dimensions by
\begin{equation}
        B_{abcd} =
        R_{aecf} R_b{}^e{}_d{}^f
        +R_{aedf} R_b{}^e{}_c{}^f
        -\frac{1}{2} g_{ab} R_{efcg} R^{ef}{}_d{}^g
        -\frac{1}{2} g_{cd} R_{aefg} R_b{}^{efg}
        +\frac{1}{8} g_{ab} g_{cd} R_{efgh} R^{efgh}.
\end{equation}
When the standard decomposition is substituted we obtain~\cite{BS}
\begin{equation}
        B_{abcd}={\cal T}_{abcd}+{\cal Q}_{abcd}+{\cal M}_{abcd}
\end{equation}
where ${\cal T}_{abcd}$ is the Bel-Robinson tensor consisting of
quadratic terms in the Weyl tensor, ${\cal M}_{abcd}$ consists of
quadratic terms in the Ricci tensor $R_{ab}$, and ${\cal Q}_{abcd}$
consists of products of Weyl and Ricci components (gravity-matter
coupling term)~\cite{sen99},
\begin{eqnarray}\label{eq:Q}
        {\cal Q}_{abcd}=&&
        \frac{1}{n-2}\Bigl(-4C^i{}_{(cd)(a}\tilde R_{b)i}
                -4C^i{}_{(ab)(c}\tilde R_{d)i}
                +2\tilde R_{ij}(C_a{}^i{}_{(c}{}^j g_{d)b}
                        -C_c{}^i{}_d{}^j g_{ab}
                        +C_b{}^i{}_{(c}{}^jg_{d)a}
                        -C_a{}^i{}_b{}^jg_{cd})\Bigr)
\nonumber\\&&
        + \frac{2}{n(n-1)}R(C_{acbd}+C_{adbc}).
\end{eqnarray}
The structure of ${\cal Q}$ (maximum of one delta) suggests that we
investigate the \emph{four} dimensional identity
$C_{[ac}{}^{[bd}\delta_{e]}^{f]}=0$; when we multiply this identity by
$\tilde R^f{}_e$ we obtain the identity
\begin{equation}\label{eq:CR}
        C_{dij[a} g_{c]b} \tilde R^{ij}
        -C_{bij[a} g_{c]d} \tilde R^{ij}
        +C_{aci[b} \tilde R_{b]}{}^i
        +C_{bdi[a} \tilde R_{c]}{}^i
        = 0.
\end{equation}
By symmetrising over the index pair $(cd)$ we get precisely the
identity
\begin{equation}
        4C^i{}_{(cd)(a}\tilde R_{b)i}
        +4C^i{}_{(ab)(c}\tilde R_{d)i}
        -2\tilde R_{ij}(C_a{}^i{}_{(c}{}^j g_{d)b}
                -C_c{}^i{}_d{}^j g_{ab}
                +C_b{}^i{}_{(c}{}^jg_{d)a}
                -C_a{}^i{}_b{}^jg_{cd})
        =0
\end{equation}
and so  the gravity-matter coupling term~\eqref{eq:Q}
simplifies, \emph{in four dimensions}, to
\begin{equation}
        {\cal Q}_{abcd}=R(C_{acbd}+C_{adbc})/6.
\end{equation}
Bonilla and Senovilla~\cite{BS} have obtained this
result, but since they were working with duals, they did not encounter the
four dimensional identity explicitly; Zund~\cite{Z} has also found 
the remarkably
simple gravity-matter coupling term, using spinors.

The possibility of the \emph{five} dimensional identity
$C_{[ab}{}^{[cd} \delta_{e}^{f}\delta_{h]}^{g]}=0$ supplying
significant simplification is obviously ruled out since, after
multiplication with one trace-free Ricci tensor, there will still be
at least one term with two deltas.
\end{example}

\begin{example}
The four dimensional identity for the Lanczos potential $L_{ab}{}^c$,
\begin{equation}
         2L^{def}g_{[a|c|}C_{b]def}
        -2L_{[a}{}^{de}C_{b]edc}
        -\frac{1}{2} L^{de}{}_c C_{deab}= 0
\end{equation}
plays an important role in the derivation of the wave equation of the
Lanczos potential~\cite{prs}. Its existence was first noted because
the spinor equivalence of the left hand side collapsed; subsequently,
it was proven by Edgar in~\cite{edgar94} by using four dimensional
duals, and also by using Lovelock's four dimensional identity
$C_{[ab}{}^{[cd} \delta_{e]}^{f]}=0$.  We can also deduce it by using
the four dimensional identity in Example~\ref{ex:lanczos2}, and then
expanding
\begin{equation}
        L_{[ab}{}^{[c}\delta_d^f\delta_{e]}^{g]} C^{de}{}_{fg} = 0.
\end{equation}
\end{example}
The remaining examples involve identitities which cannot be deduced 
\emph{directly} from
Lovelock's identities.

\begin{example}\label{ex:ricciid}
In~\cite{bonanos98}, Bonanos demonstrated that a complicated tensor
\begin{eqnarray}
    \chi'_{abcd}
     = &&\tilde R_{ac}\tilde R_b{}^m\tilde R_{md}
    + \tilde R_{bd}\tilde R_a{}^m\tilde R_{mc}
    - \tilde R_{ad}\tilde R_b{}^m\tilde R_{mc}
    - \tilde R_{bc}\tilde R_a{}^m\tilde R_{md}
    + \tilde R_{a}{}^m\tilde R_m{}^n\tilde R_{nc}g_{bd}
    + \tilde R_{b}{}^m\tilde R_m{}^n\tilde R_{nd}g_{ac}
\nonumber \\&&
    - \tilde R_{a}{}^m\tilde R_m{}^n\tilde R_{nd}g_{bc}
    - \tilde R_{b}{}^m\tilde R_m{}^n\tilde R_{nc}g_{ad}
\nonumber \\&&
    - \frac{1}{2}(\tilde R_{mn}\tilde R^{mn})
         (\tilde R_{ac}g_{bd}+\tilde R_{bd}g_{ac}
         - \tilde R_{ad}g_{bc}
         - \tilde R_{bc}g_{ad})
    - \frac{1}{3}(\tilde R_{mn}\tilde R^{mr}\tilde R^n{}_r)
          (g_{ac}g_{bd}-g_{ad}g_{bc})
\end{eqnarray}
of third order in the trace-free Ricci tensor $\tilde R_{ab}$, which
had been used in the study of Riemann invariants in four
dimensions~\cite{zakhary97}, was in fact, surprisingly, identically
zero.

In~\cite{edgar99} it was demonstrated how this result could be seen as
an {\it indirect} consequence of Lovelock's results~\cite{lovelock70}.
By applying Theorem \ref{th:lovelock2} to the trace-free Plebanski
tensor $P_{ab}{}^{cd}$ which is the 'square' of the Ricci tensor given
by
\begin{equation}\label{eq:ricciid2}
        P_{ab}{}^{cd}
        =
        2 \tilde R_{[a}{}^{[c}\tilde R_{b]}{}^{d]}
        +2 \tilde R_{[a}{}^{i}\tilde R_{|i|}{}^{[c}\delta_{b]}^{d]}
        - \frac{1}{3} \tilde R_{j}{}^{i}\tilde R_{i}{}^{j}
                \delta^c_{[a} \delta^{d}_{b]}
\end{equation}
the following identity of second order in the trace-free Ricci tensor was
obtained
\begin{equation}\label{eq:ricciid3}
        2 \tilde R_{[a}{}^{[c}\tilde R_{b}{}^{d}\delta_{e]}^{f]}
        +2 \tilde R_{[a}{}^{i}\tilde R_{|i|}{}^{[c}\delta_b^d\delta_{e]}^{f]}
        -\frac{1}{3} \tilde R_{j}{}^{i}\tilde R_{i}{}^{j}
                \delta^c_{[a}\delta^d_b\delta^f_{e]}
        =0.
\end{equation}
It was then shown in~\cite{edgar99} that the identically zero tensor
$\chi'_{abcd}$ found by Bonanos was just a direct consequence of
multiplying the identity~\eqref{eq:ricciid3} by $\tilde R_f{}^e$.

But we now have the complete picture.  The basic identity in four
dimensions for the trace-free Ricci tensor is the first order
identity~\eqref{eq:Ricci1}
\begin{equation}\label{eq:ricciid}
        \tilde R_{[a}{}^{[c} \delta_b^d\delta_{e}^{f}\delta_{g]}^{h]}
        = 0.
\end{equation}
By successive multiplications by the trace-free Ricci tensor we
obtain, first of all, the identity (\ref{eq:ricciid3}) which is second
order in the trace-free Ricci tensor; subsequently by multiplying the
left hand side of (\ref{eq:ricciid}) by $\tilde R_f{}^e \tilde
R_h{}^g$ we obtain the third order identity which is $\chi'_{abcd}$
identically zero, and finally the fourth order identity, which is the
Cayley-Hamilton theorem for the matrix representation of $\tilde
R_a{}^b$ in four dimensions, as shown in~\cite{edgar99}.
\end{example}

\begin{example}\label{ex:sen1}
The Bel-Robinson tensor~\cite{Rob} is given in $n$ dimensions by
\begin{equation}\label{eq:BR}
        {\cal T}_{abcd} =
        C_{aecf} C_b{}^e{}_d{}^f
        +C_{aedf} C_b{}^e{}_c{}^f
        -\frac{1}{2} g_{ab} C_{efcg} C^{ef}{}_d{}^g
        -\frac{1}{2} g_{cd} C_{aefg} C_b{}^{efg}
        +\frac{1}{8} g_{ab} g_{cd} C_{efgh} C^{efgh}.
\end{equation}
It is obviously symmetric over the first and last pair of indices, but in
order to investigate its symmetry over \emph{all} indices we need to examine,
\begin{equation}\label{eq:sen1}
        {\cal T}_{a[bc]d}
        = 
        \frac{1}{4} C_{adef} C_{bc}{}^{ef}
        -C_{eaf[b} C_{c]}{}^e{}_d{}^f
        -C_{fge[a} g_{d][b} C_{c]}{}^{efg}
        +\frac{1}{8} g_{a[b} g_{c]d} C_{efgh} C^{efgh}.
\end{equation}
Its structure (maximum of two deltas) suggests that we investigate the
five dimensional identity $C_{[bc}{}^{[ad}
\delta_{e}^{g}\delta_{f]}^{h]}=0$; by multiplying with $C_{gh}{}^{ef}$
we obtain precisely the right hand side of~\eqref{eq:sen1}. So the
Bel-Robinson tensor is completely symmetric in both four and five
dimensions, but of course this calculation does not give us any
information about higher dimension.  (Although we
know that $C_{[ab}{}^{[cd} \delta_{g}^{e}\delta_{h]}^{f]}=0$ is
\emph{not} an identity in higher dimensions, we are considering the
more complicated expression~\eqref{eq:sen1}.)  However, by taking the
non trivial double trace on~\eqref{eq:sen1} we obtain
\begin{equation}
        {\cal T}^a{}_{[ab]}{}^b
        = \frac{(n-4)(n-5)}{16} C_{abcd} C^{abcd}
\end{equation}
which shows that the dimensions four and five are both necessary and sufficient
conditions for the Bel-Robinson tensor to be completely symmetric.

The fact that ${\cal T}_{abcd}$ is completely symmetric, in, and only
in, dimensions four and five was confirmed by Senovilla~\cite{sen99}
from its definition in terms of duals, but in a less direct manner,
where each dimension was considered seperately.
\end{example}

\begin{example}\label{ex:LanE}
The Bel-Robinson tensor (constructed from the Weyl tensor), discussed
in the last example, has the wrong dimension for energy, so
Roberts~\cite{Robs} has proposed instead using an analogous
construction with the Lanczos potential of the Weyl tensor which has
the correct dimensions for energy. He has suggested looking at the
most general expression quadratic in the Lanczos potential, but if
instead we use Senovilla's definition of {\bf super-energy
tensor}~\cite{sen99} we find the super-energy tensor associated with
the Lanczos potential $L_{ab}{}^c$ (a trace-free (2,1)-form) is given
in $n$ dimensions by
\begin{equation}\label{eq:BRL}
        {\cal T}^L_{abcd} =
        L_{aec} L_b{}^e{}_d{}
        +L_{aed} L_b{}^e{}_c{}
        -\frac{1}{2} g_{ab} L_{efc} L^{ef}{}_d{}
        - g_{cd} L_{aef} L_b{}^{ef}
        +\frac{1}{4} g_{ab} g_{cd} L_{efg} L^{efg}.
\end{equation}
It is obviously symmetric over the first and last pair of indices
respectively.  It is not symmetric over all indices which can be shown
by choosing a local orthonormalised basis and
$L_{131}=-L_{311}=-L_{232}=L_{322}=1$, all others zero in $n$
dimensions.  Then ${\cal T}^L_{1[12]2} = \pm\frac{1}{2}$ (the sign
depends on the signature).

The Bel-Robinson tensor in the previous example had the pairwise
symmetry ${\cal T}_{abcd}={\cal T}_{cdab}$.  That symmetry can be
imposed on the super-energy tensor for the Lanczos potential by examining
\def\tsym{T^L{}}
\begin{equation}
        \tsym_{abcd} =
        \frac{1}{2}{\cal T}^L_{abcd}
        +\frac{1}{2}{\cal T}^L_{cdab}
\end{equation}
instead.  In order to investigate whether this makes it symmetric over
\emph{all} indices we examine
\begin{equation}\label{eq:Tasym}
        \tsym^{a}{}_{[bc]}{}^{d}
        =
        \tsym^{[a}{}_{[bc]}{}^{d]}
        =
        \frac{1}{4} L_{bce} L^{ade}
        - L^{[a}{}_{e[b} L_{c]}{}^{|e|d]}
        - \frac{1}{2} \delta^{[a}_{[b} L_{|ef|c]} L^{|ef|d]}
        - \delta^{[a}_{[b} L_{c]ef} L^{d]ef}
        + \frac{1}{4} \delta^a_{[b} \delta^d_{c]} L_{efg} L^{efg}.
\end{equation}

Its structure (maximum of two deltas) suggests that we investigate the
four dimensional identity $L_{[ef}{}^{[g} \delta_b^a
\delta_{c]}^{d]}=0$.  By multiplying with $L^{ef}{}_{g}$ we obtain the
right hand side of~\eqref{eq:Tasym}, so $\tsym_{abcd}$ is completely
symmetric in four dimensions and lower.  It is, however, necessary to
use the additional symmetry of the Lanczos potential $L_{[abc]}=0$ so
this result does not hold for any $(2,1)$-form.

By taking the non trivial double trace of~\eqref{eq:Tasym} we obtain
(by using $L_{[abc]}=0$ again)
\begin{equation}
        \tsym^a{}_{[ab]}{}^b = \frac{(n-4)(n-3)}{8} L_{abc} L^{abc}
\end{equation}
which shows that the dimension being four or less is both a necessary
and sufficient condition for the symmetrised super-energy tensor
$\tsym_{abcd}$ for the Lanczos potential to be completely symmetric.

Although we can follow Senovilla's construction for the super energy
tensor from any $(2,1)$-form $L_{ab}{}^c$ in $n$ dimensions, we
should point out that the Lanczos potential for the Weyl tensor is
unlikely to exist generally in dimensions above four~\cite{eh00}.
\end{example}

\section{Summary and Discussion}

Of course,  Lovelock's identities and our generalisations found in
this paper are not really 'new'   since they are just simple and direct
specialisations of those fundamental identities found by antisymmetrising over
$n+1$ indices in
$n$ dimensions; furthermore, they are not complete alternatives for those
fundamental identities, since, being their traces, they contain less
information. Rather, the significance of these identities is that
they highlight the fact that there exist heavily disguised versions of these
fundamental identities when trace-free and antisymmetry properties are also
introduced; and since we often are dealing with tensors with these explicit
properties it is often the  specialisations of the fundamental
identities which are relevant in practical applications.

The use of dimensionally dependent identities is a powerful method
which has been largely overlooked, perhaps because of its
simplicity. A very striking example of this is to be seen in the
algebraic Rainich condition for the electromagnetic energy tensor in
four dimensions. Over the past 75 years this condition has been
obtained by a variety of very different methods: Rainich~\cite{R} used
invariant planes, while others have used duality rotations~\cite{Wh},
complex duals~\cite{Wi,GN} and complicated matrix manipulation based
on the Cayley-Hamilton theorem~\cite{abs}; in spinors, a simple direct
calculation has been given~\cite{penrose1}. In fact Rainich's result
was one of the motivations for Lovelock's work~\cite{lovelock70} and
although he rederived the result using an explicit four dimensional
identity,~\cite{lovelock67,lovelock70} it is still a somewhat
roundabout and contrived calculation.  We shall now show that the
result can be viewed as simply the four dimensional Cayley-Hamilton
theorem when considered as a dimensionally dependent identity, or
equivalently a trivial application of our basic result.

\begin{example}\label{ex:MaxwellE}
The four dimensional algebraic Rainich identity associated with the
energy momentum tensor $T^a{}_b= \Ful ac \Ful cb - \frac{1}{4} \delta^a_b
\Ful cd \Ful dc$ of an electromagnetic field $F_{ab}$ is given by
\begin{equation}
        T^a{}_c T^c{}_b = \frac{1}{4} \delta^a_b T^c{}_d T^d{}_c.
\end{equation}
When written out
in full this identity is
\begin{equation}\label{eq:Max}
        \Ful ac \Ful cd \Ful de \Ful eb
        -\frac{1}{2} \Ful ac \Ful cb \Ful de \Ful ed
        -\frac{1}{4} \delta^a_b \Ful cd \Ful de \Ful ef \Ful fc
        +\frac{1}{8} \delta^a_b \Ful cd \Ful dc \Ful ef \Ful fe
        =0.
\end{equation}
But this is simply the Cayley-Hamilton theorem for $\Ful ab$ as given in
Example~\ref{ex:CH},
\begin{equation}
        F^c{}_{[c}F^d{}_{d}F^e{}_eF^f{}_f \delta^a_{b]}
        =0
\end{equation}
when specialised to antisymmetric $F_{ab}$.  Equivalently the structure
immmediately suggests specialising the four dimensional identity for
trace-free $(1,1)$-forms
$F^{[f}{}_{[c}\delta_d^g\delta_e^h\delta_{b]}^{a]}=0$ given
in~\eqref{eq:Ricci1} to antisymmetric forms, from which we obtain the
identity
\begin{equation}
        F^{[f}{}_{[c}\delta_d^g\delta_e^h\delta_{b]}^{a]}
        F^c{}_f F^d{}_g F^e{}_h
        =0
\end{equation}
which, when expanded and specialised to antisymmetric $F_{ab}$, is
identical to~\eqref{eq:Max}.

A more direct way to obtain~\eqref{eq:Max} is to expand
\begin{equation}
        F^{cd}F^{ef}F_{[cd}F_{ef}\delta^{a}_{b]}=0.
\end{equation}
\end{example}

In this particular example, as in the others in the last section, we
have been able to focus directly on the fundamental dimensionally
dependent identity underlying the result.  By identifying this
underlying identity in our various applications to four dimensions in
this paper, we are in a position to explore directly the possibility
of generalisations to other dimensions and to other forms.  We shall
present such generalisations in a subsequent paper.

Of course, it is possible to establish results peculiar to four dimensions
without having to deal explicitly with dimensionally dependent 
identities; this is most
easily done by using spinors, where the dimension is built into the 
formalism. Indeed,
it was  the apparent discrepency between results in spinors and tensors which
originally gave the clue to the existence of a number of these four dimensional
identities. If we were only dealing with four dimensional spaces with Lorentz
signature, then spinors would be the more efficient formalism to use; 
on the other
hand, the advantage of the tensor formalism, is that once we have 
identified the
underlying four dimensional tensor identity, generalisations to other 
dimensions can be
sought.

In a similar way, when explicit four dimensional duals are used, then
the dimension can also be built in by some of the identities satisfied
by duals. For instance, the key identities associated with duals in
work on super-energy tensors by Senovilla (equations (2) and (3)
in~\cite{sen99}) have dimension $n$ built in explicitly; so these are
dimensionally dependent identities constructed from identities of the
form
\begin{equation}
        T^{\cal {A}}_{[a_1}\eta_{ a_2 a_3 ... a_{n+1} ]} = 0.
\end{equation}
In this context, it is significant that the four dimensional
identity~\eqref{eq:Weylsquare} is obtained in~\cite{sen99} essentially
by taking double duals.

As emphasised in the last section,  important applications of these 
results will be to
Riemann tensors. Of
course the Riemann tensor has additional symmetries to those of a
$(2,2)$-form;
these additional symmetries have not been explicitly considered in 
this paper, but
would need to be considered explicitly in an exhaustive treatment of 
invariants of the
Riemann tensor.

In conclusion, we emphasise that when examining what actually was required
in the
proofs of the theorems in  Section 4 one discovers that very
little structure was needed.  Firstly, the results hold pointwise so
the manifold structure is not needed.  Secondly, there was no raising
or lowering of indices so no assumption on what to use for that is
needed; indeed not even the existence of such operations is needed.
Thirdly, no metric was used.  This means that those theorems can be
applied to objects other than tensors: e.g. spinors when $n=2$;
matrices when the number of indices on the objects are appropriate; or
other indexed objects such as Christoffel symbols or tensors in
coordinate index notation or tetrad index notation.

\begin{acknowledgments}
SBE is grateful for support from the Swedish Natural Science Research
Council.
\end{acknowledgments}

\end{document}